\documentclass[journal]{IEEEtran}
\usepackage{ifpdf}
\usepackage{algorithmic}
\usepackage{cuted}
\usepackage{flushend}
\usepackage{algorithm}
\ifCLASSINFOpdf
   \usepackage[pdftex]{graphicx}

\else
 \usepackage[dvips]{graphicx}
\fi

\ifCLASSOPTIONcompsoc
    \usepackage[caption=false, font=normalsize, labelfont=sf, textfont=sf]{subfig}
\else
\usepackage[caption=false, font=footnotesize]{subfig}
\fi
\usepackage[cmex10]{amsmath}
\usepackage{epstopdf}
\usepackage{array} 
\usepackage{flushend}
\usepackage{balance}
\usepackage{eqparbox}
\usepackage[noadjust]{cite}
\usepackage{float}

\usepackage{stfloats}
\usepackage{url}
\usepackage{amsthm}

\usepackage{amssymb}
\begin{document}

\title{RIS-Aided Physical Layer Security Improvement in Underlay Cognitive Radio Networks}

\setlength{\columnsep}{0.21 in}

\author{Majid H. Khoshafa,~\IEEEmembership{Member,~IEEE,}
        Telex M. N. Ngatchede,~\IEEEmembership{Senior Member,~IEEE,} \\
        and~Mohamed H. Ahmed,~\IEEEmembership{Senior Member,~IEEE,}
\thanks{M. H. Khoshafa is with the Department of Electrical and Computer Engineering, Queen's University, Kingston, Canada (e-mail: mhk5@queensu.ca)}
\thanks{T. M. N. Ngatched is with the Department of Electrical and Computer Engineering, Memorial University of Newfoundland, St. John's, Canada (e-mail: tngatched@grenfell.mun.ca)}
\thanks{M. H. Ahmed is with the School of Electrical Engineering and Computer Science, University of Ottawa, Ottawa, Canada (e-mail: mahme3@uottawa.ca)}
}
\maketitle

\begin{abstract}
 In this paper, a reconfigurable intelligent surface (RIS)-aided underlay cognitive radio network is investigated. An RIS is utilized to improve the secondary network (SN) reliability and robustness while simultaneously increasing the physical layer security of the primary network (PN). Toward this end, closed-form expressions for the SN outage probability, PN
secrecy outage probability, and PN probability of non-zero secrecy capacity are derived. To increase the eavesdropping signals of the PN, the eavesdropper uses two combining techniques, namely maximal ratio combining and selection combining. Furthermore, the advantages of the proposed system model are verified through numerical and simulation results.

\end{abstract}
\begin{IEEEkeywords}
Reconfigurable intelligent surfaces, cognitive radio network, physical layer security.
\end{IEEEkeywords}
\IEEEpeerreviewmaketitle
\section{Introduction}
\IEEEPARstart{R}{}ECONFIGURABLE Intelligent Surfaces (RISs) are attracting much consideration as a leading technology to achieve intelligent wireless channels environment for the next generation networks \cite{basar2019wireless}. RISs are planar surfaces of electromagnetic (EM) material comprising a large number of cheap passive reflecting elements. A microcontroller controls each element to alter the amplitude and phase of the reflected signal. The RIS technology has many advantages, including the ability to change transmission environments into intelligent ones, enhancing the quality of the received signals at the destination,  reducing the power consumption compared with other technologies, increasing the physical layer security (PLS), and alleviating the undesired interference  \cite{wu2019towards, wu2021intelligent, RISKhoshafa, 9530403}. {Passive RISs prototypes were assembled in \cite{fara2022prototype,  trichopoulos2022design,  dai2020reconfigurable} to acquire more practical and precise results regarding the actual performance of RISs-aided systems by taking experimental measurements.}

With the envision new technologies and utilizing higher frequency bands, secure communications are significant in the sixth generation (6G) wireless networks, where new security challenges arise \cite{chorti2022context}. {Present research contributions have established RISs as cutting-edge technology, with promising research directions toward the 6G. To take things further, integrating RISs with emerging communication technologies results in higher performance gains that can be achieved \cite{basharat2021reconfigurable}}.  The PLS, initially investigated by Wyner \cite{wyner1975wire},  has evolved as an attractive technique for improving the cellular network's secrecy performance against signal leakage. In this respect,  PLS utilizes the natural properties and characteristics of wireless communication channels and noise to secure data transmission by limiting the amount of data that can be leaked at the bit level by eavesdroppers. Thanks to their distinctive characteristics, which enable them to control the transmission environment, RISs can be utilized to eliminate interference and improve the received signal without using active elements. In this respect, the RIS technology has been recently utilized to improve the PLS of wireless communication system \cite{9652031,guan2020intelligent,RISKhoshafa}. To guarantee a secure transmission, the RIS was deployed near the eavesdropper to cancel out the eavesdropping signal received by the eavesdropper \cite{zhang2021improving}, which can actually decrease the information leakage to enhance the PLS of the wireless network.

On the other hand, mobile  wireless communication has experienced rapid development in data traffic due to the dramatic growth of smart devices. According to Cisco, the average number of mobiles per capita will be 3.6 by 2023 \cite{cisco}, leading to an enormous demand for radio spectrum resources, including bandwidth and energy. Consequently, spectral and energy efficiency are two crucial principles for designing future wireless networks. Cognitive radio (CR) has been introduced as an efficient technique to improve spectral efficiency. In CR networks, the spectrum can be shared by two different networks, the primary network (PN) and the secondary network (SN), provided that the interference produced by the SN to the PN is controlled by interference constraint. {The authors in \cite{khoshafa2021performance} studied the PLS of energy harvesting for CR networks using the cooperative relaying technique. 
} A common technique is to employ beamforming to improve the performance of the SN while guaranteeing that the interference power received by the PN users is below the predefined interference limit. Nevertheless, the beamforming gain is restricted when the link between the SN transmitter and SN receiver is weak due to severe attenuation. To address such a problem, an RIS can be deployed to improve the performance of the SN while enhancing the secrecy rate of the PN \cite{pan2021reconfigurable}. In \cite{zhang2020intelligent}, the RIS technology has been employed to aid data transmission in CR networks. The authors in \cite{yuan2020intelligent} proposed an RIS-assisted CR network to enhance the SN's achievable rate. In this work, we propose a secure RISs-aided underlay CR network. To the best of our knowledge, there is no previous work studying the advantage of the RIS technology to secure CR network. Furthermore, the influence of the RIS technology on the PN secrecy capacity is investigated. The main contributions of this paper can be summarized as follows:
\begin{figure}[t]
\centering
\includegraphics[width=3.2in]{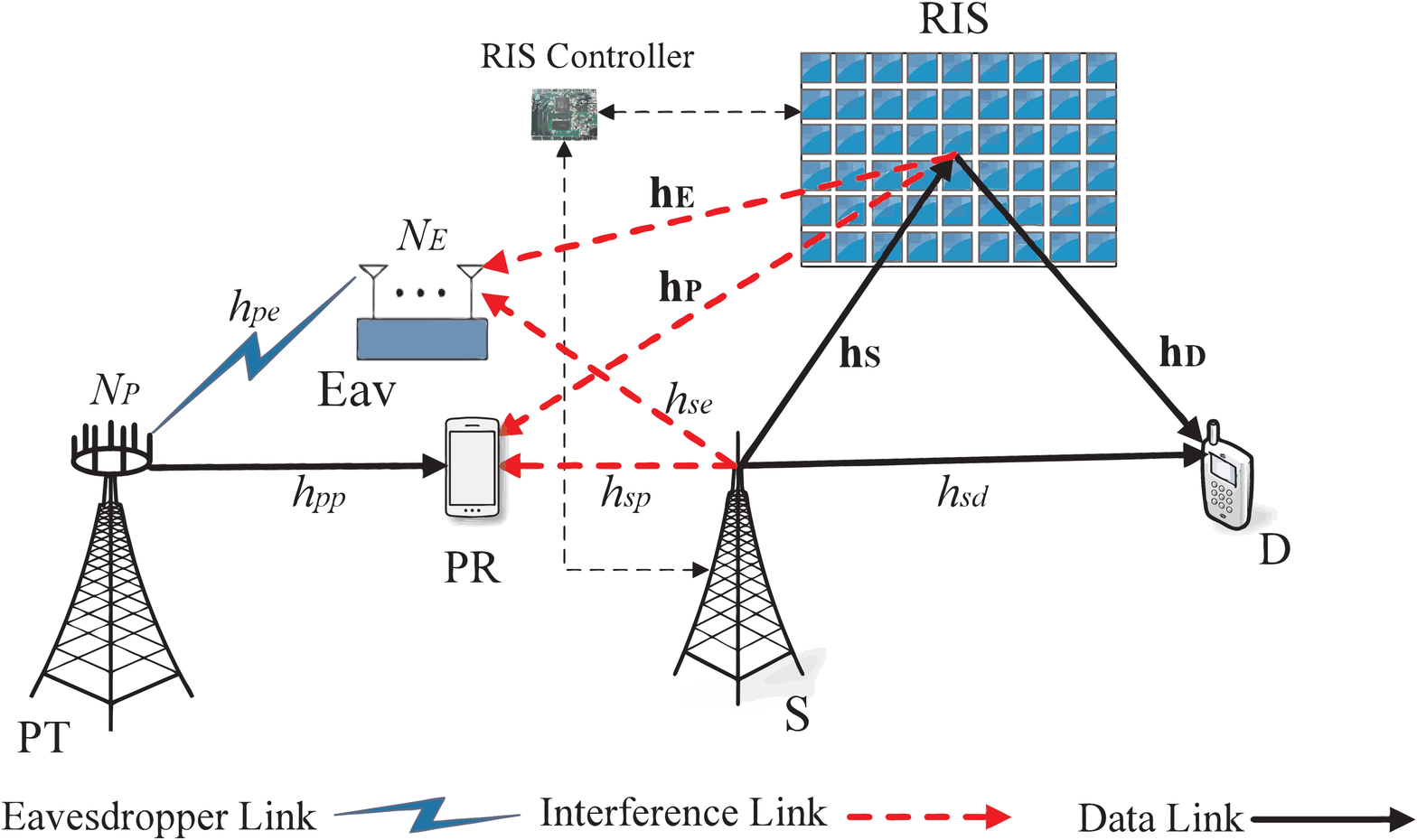}
\caption{{System Model.}}
\label{Sys}
\end{figure}
\begin{itemize}   
\item The RIS technology is introduced to improve the reliability of the SN, while concurrently increasing the physical layer security of the PN.

\item To compensate for the spectrum sharing, the RIS technology is utilized as a friendly jammer to ensure a high-secrecy performance for the PN, consequently enabling a win-win situation between the two networks, i.e., security provisioning for the PN and high reliability and robustness for the SN.
\item The SN outage probability is studied, and a novel analytical expression is derived. Besides, closed-form expressions for the secrecy outage probability (SOP) of the PN are also derived, considering two combining techniques, namely maximal ratio combining (MRC) and selection combining (SC), which are employed by an eavesdropper.
\item Asymptotic analysis is provided for the SOP of the PN. Moreover, the benefits of the proposed system model are confirmed through numerical and simulation results.
\end{itemize}

\begin{table}
\centering
\caption{{Table of Symbols}}
\label{table}
\setlength{\tabcolsep}{3pt}
\begin{tabular}{|p{25pt}|p{180pt}|}
\hline
{Symbol}& 
{Description}\vspace{2mm} \\
\hline
S& 
Secondary transmitter \\
D& 
Secondary receiver\\
PT& 
Primary transmitter\\
PR& 
Primary receiver\\
Eav& 
Eavesdropper\\
$N_P$& 
Number of antennas at PT\\
$N_E$& 
Number of antennas at Eav\\
$N$& 
Number of reflecting elements at RIS\\
$h_{ab}$& 
Channel coefficient of $ab$ link\\
$d_{ab}$& 
Euclidean distance between $ab$\\
$n_{a}$& 
AWGN node at $a$\\
$\sigma^2_a$& 
AWGN variance at node $a$\\
$y_a$& 
Received signal at node $a$\\
$x_s$& 
SN transmitted signal \\
$x_p$& 
PN transmitted signal \\
$P_\textup{S}$& 
SN transmitted power\\
$P_\textup{P}$& 
PN transmitted power\\
$d_o$& 
Reference distance\\
$\phi_{i}$& 
Phase coefficient of the $i^{th}$ element of the RIS\\
$\theta _{i}$& 
Residual phase errors affecting the PR\\
$\psi_{i}$& 
Residual phase errors affecting the Eav\\
$\gamma_a$& 
SINR at node $a$\\
$\bar{\gamma}_a$ & 
Average SNR at node $a$\\
$\eta$& 
Path loss exponent\\
$f_{X}(\cdot)$& 
PDF of random variable $X$\\
$F_{X}(\cdot)$& 
CDF of random variable $X$\\
$F^{ \infty }_{X}(\cdot)$& 
Asymptotic CDF of random variable $X$\\
$\mathcal{Q}$& 
Threshold of interference temperature\\
$\mathcal{C}_\textup{S}$& 
PN secrecy capacity\\
$\mathcal{C}_\textup{P}$& 
PN capacity\\
$\mathcal{C}_\textup{E}$& 
Eav capacity\\
$\mathcal{R}_d$& 
SN achievable data rate\\
$\mathcal{R}_s$& 
PN target secrecy rate\\
$\textrm{Pr}(\cdot)$& 
Probability of an event\\
SOP& 
Secrecy outage probability\\
$P_{out}$& 
SN outage probability\\
$\textup{SOP}^{ \infty }$& 
Asymptotic SOP\\
$\mathcal{G}_a$& 
Secrecy diversity order\\
$\mathcal{G}_d$& 
Secrecy array gain\\
$\mathcal{O}(\cdot)$& 
Higher order term\\
$\beta$& 
$2^{\mathcal{R}_s}$\\
$\alpha$& 
$\beta-1$ \\
\hline
\end{tabular}
\label{tab1}
\end{table}
%


\section{System Model}
\label{S01}
A proposed RISs-aided underlay CR system, including a license-holding PN and an unlicensed SN, is considered as shown in Fig. \ref{Sys}. Specifically, the SN comprises a secondary transmitter (S) and a secondary receiver (D), each equipped with a single antenna, while the PN comprises a primary transmitter (PT) and a single-antenna primary receiver (PR). The PT is equipped with $N_P\geq1$ antennas. In addition, an eavesdropper (Eav), equipped with $N_E\geq1$ antennas, intends to overhear the PN's data streams. Therefore, an RIS, made of $N$ reflecting elements, is utilized to enhance the achievable secrecy rate of the PN by interfering with the eavesdropping signals at Eav while improving the transmission conditions of the SN.  {It is worth mentioning that a field-programmable gate array (FPGA) can be utilized as a controller to achieve adjustable control of the RIS in practice, which often communicates and coordinates with other network elements (e.g., BS and users) via dedicated connections  \cite{shi2022wireless}.} It is assumed that the channel state information (CSI) of all channels employed in the system is known\footnote{The traditional pilot signaling techniques can be utilized to estimate the CSI of the legitimate transmission links \cite{basar2019wireless,  zhang2020intelligent, guan2020intelligent, 76543, 9291402} }. As we consider a passive Eav, its CSI is unknown at both the RIS and the PT.   

The channel coefficients for the
$\textup{PT} \to \textup{PR} ,\:
\textup{PT} \to \textup{Eav} ,\:
\textup{RIS} \to
\textup{Eav},\:
\textup{RIS} \to
\textup{PR},\:
\textup{S} \to \textup{Eav},\: \textup{S} \to \textup{PR},\: \textup{S} \to \textup{D} ,\:\textup{S} \to \textup{RIS},$ and $\textup{RIS} \to \textup{D}$ links are expressed as $h_{pp},\: h_{pe},\: \textbf{\textrm{h}}_\textbf{\textrm{E}}\footnote{ The bold font is used to indicate vectors.},\: \textbf{\textrm{h}}_\textbf{\textrm{P}},\:h_{se}, \:h_{sp},\: h_{sd},\: \textbf{\textrm{h}}_\textbf{\textrm{S}},$ and $\textbf{\textrm{h}}_\textbf{\textrm{D}},$ respectively. The above channel coefficients are assumed to undergo Rayleigh fading\footnote{Since the transmission links experience blockages and the RIS's location cannot be optimized to guarantee reliable line-of-sight links, similar to \cite{ guan2020intelligent, 76543, 9291402}, Rayleigh fading environment is assumed in this work. 
}. {To elaborate, $\textbf{\textrm{h}}_\textbf{\textrm{S}}\in \mathbb{C}^{N\times 1},\: \textbf{\textrm{h}}_\textbf{\textrm{D}}\in \mathbb{C}^{1\times N},\: \textbf{\textrm{h}}_\textbf{\textrm{E}}\in \mathbb{C}^{1\times N},$ and $
\textbf{\textrm{h}}_\textbf{\textrm{P}}\in \mathbb{C}^{1\times N}$ denote the channel vector between the SN transmitter and RIS, RIS and SN receiver, RIS and eavesdropper, and RIS and PN receiver, respectively.}  Moreover, the Euclidean
distances between $\textup{S} \to \textup{RIS} ,\: \textup{RIS} \to \textup{D},\:
\textup{S} \to \textup{D} ,\:
\textup{PT} \to \textup{PR} ,\:
\textup{PT} \to \textup{Eav} ,\:
\textup{RIS} \to
\textup{Eav},\:
\textup{RIS} \to
\textup{PR},\:
\textup{S} \to \textup{Eav},$ and $ \textup{S} \to \textup{PR}$ links are denoted as $d_{sr},\:d_{rd},\: d_{sd},\: d_{pp},\: d_{pe},\: d_{re},\: d_{rp},\:d_{se},$ and $\:d_{sp},$ respectively. In addition, $n_{\varrho  }$ is the additive white
Gaussian noise (AWGN) at D, PR, and E, respectively, where $\varrho  \in \left \{ d, p, e \right \}$, with zero mean and variance  $\sigma^{2}_{\varrho  }$. Consequently, the received signal at D can be written as
\begin{equation}\begin{split}    y_{\textup{D}} =\sqrt{P_\textup{S}}\,x_s&\left [ \left ( \frac{d_{sr}\,  d_{rd}}{d_{o}^{2}} \right )^{-\frac{\eta }{2}}\sum_{i=1}^{N} h_{s_i}\, h_{d_i}\, e^{j\phi_{i}}\right.\\&\left.+h_{sd}\left ( \frac{d_{sd}}{d_{o}} \right )^{-\frac{\eta }{2}}  \right ]+n_d,  \end{split}\end{equation}
where $x_s$ is the SN transmitted signal, $P_\textup{S}$ denotes the SN transmitted power, $d_{o}$ is a reference distance, and $\eta$ is the path loss exponent. In addition, $h_{s_i}$ and $h_{d_i}$ are complex Gaussian random variables (RV) with a zero mean and unit variance, and $\phi_{i}$ is the alterable phase coefficient of the $i^{th}$ element of the RIS. Moreover, it is assumed that the PT position is distant from the RIS and D and, therefore, does not impose any real interference. Consequently, the interference at the RIS and D from PT is negligible; this is a well-known assumption that is widely used in the literature  \cite{lee2010outage, lei2017secrecy}. Furthermore,  the phases of the channels $h_{s_i}$ and $h_{d_i}$ are assumed to be perfectly known at the RIS. Hence, the optimal phase shift is selected to maximize the instantaneous signal-to-noise ratio (SNR) at D \cite{basar2019wireless, bjornson2019intelligent}.  
Besides, the reflected gain of the $i^{th}$ reflecting element is assumed to equal to one \cite{guan2020intelligent, basar2019wireless}. Thus, the received signal at the $\textup{PR}$ can be written as
\begin{equation}\begin{split}   y_{\textup{P}} =&\sqrt{P_\textup{P}}\left ( \frac{d_{pp}}{d_{o}} \right )^{-\frac{\eta }{2}} \,h_{pp}\,x_p+\sqrt{P_\textup{S}}\,\, x_{s}\left [ \left ( \frac{d_{sr}\,  d_{rp}}{d_{o}^{2}} \right )^{-\frac{\eta }{2}} \right.\\&\left.\times\sum_{i=1}^{N} h_{s_i}\, h_{p_i}\, e^{j\theta_{i}}+h_{sp}\left ( \frac{d_{sp}}{d_{o}} \right )^{-\frac{\eta }{2}} \right ]+n_p,   \label{yp}    \end{split}\end{equation}
where $x_p$ is the PN transmitted signal, $P_\textup{P}$ denotes the PN transmitted power, and $\theta _{i}$ is the residual phase errors affecting the PR. In a similar way, the wiretapped signal at Eav can be written as
\begin{equation}\begin{split}    y_{\textup{E}} =&\sqrt{P_\textup{P}}\left ( \frac{d_{pe}}{d_{o}} \right )^{-\frac{\eta }{2}}\,h_{pe}\,x_p+\sqrt{P_\textup{S}}\,x_{s} \left [\left ( \frac{d_{sr}\,  d_{re}}{d_{o}^{2}} \right )^{-\frac{\eta }{2}}\right.\\&\left.\times\sum_{i=1}^{N} h_{s_i}\, h_{e_i}\, e^{j\psi_{i}} +h_{se}\left ( \frac{d_{se}}{d_{o}} \right )^{-\frac{\eta }{2}}\right ]+n_e,\end{split}\end{equation}
where $h_{e_i}$ is the channel coefficient between Eav and the $i^{th}$ reflecting element of the RIS and $\psi_{i}$ is the residual phase errors affecting the Eav. The instantaneous SNR at D, $\gamma_{D}$, is given by 
\begin{equation}
    \gamma_{\textup{D}}  
=\frac{P_\textup{S}}{\sigma_d^2}\left |\left ( \frac{d_{sr}  d_{re}}{d_{o}^{2}} \right )^{-\frac{\eta}{2}}\sum_{i=1}^{N} h_{s_i} h_{d_i} e^{j\phi_{i}}+  h_{sd}\left ( \frac{d_{sd}}{d_{o}} \right )^{-\frac{\eta}{2}} \right |^{2}.
    \label{S11}
\end{equation}

Moreover, the instantaneous signal-to-interference-and-noise ratio (SINR) at PR and Eav, denoted as $\gamma_P$ and $\gamma_E$, respectively, are given by  

\begin{equation}
    \begin{split}
        \hspace {-0.5pc}\gamma_{\textup{P}} \hspace {-0.2pc}=\hspace {-0.2pc}\frac{P_\textup{P}\left ( \frac{d_{pp}}{d_{o}^{2}} \right )^{-\eta}\left | h_{pp}\right |^{2}}{P_\textup{S}\hspace {-0.2pc}\left |\left ( \frac{d_{sr}  d_{rp}}{d_{o}^{2}} \right )^{-\frac{\eta}{2}}\sum_{i=1}^{N} h_{s_i} h_{p_i}e^{j\theta_i }\hspace {-0.2pc}+\hspace {-0.2pc}  \left ( \frac{d_{sp}}{d_{o}} \right )^{-\frac{\eta}{2}}h_{sp} \right |^{2}\hspace {-0.3pc}+\sigma_p^2}\hspace {-0.2pc},
        \label{YPP25}
            \end{split}
\end{equation}

and

\begin{equation}
    \begin{split}
        \hspace {-0.2pc}\gamma_{\textup{E}} &\hspace {-0.2pc}=\hspace {-0.3pc}\frac{P_\textup{P}\left ( \frac{d_{sd}}{d_{o}} \right )^{-\eta}\left | h_{pe}\right |^{2}}{P_\textup{S}\left |\left ( \frac{d_{sr}  d_{re}}{d_{o}^{2}} \right )^{-\frac{\eta}{2}}\hspace {-0.2pc}\sum_{i=1}^{N} h_{s_i} h_{e_i}e^{j\psi_i }+\left ( \frac{d_{se}}{d_{o}} \right )^{-\frac{\eta}{2}}h_{se}  \right |^{2}\hspace{-0.3pc}+\hspace {-0.3pc}\sigma_e^{2}},
        \label{YYE}
    \end{split}
\end{equation}
which can be rewritten as
\begin{align}   \begin{split}       \gamma_{\textup{E}} =\frac{\Psi_\textup{PE}}{\Psi_\textup{E}+1},\label{c756}    \end{split}\end{align}
where 
\begin{align*}   \begin{split}       \Psi_\textup{E}=\left |\Lambda_{1} \sum_{i=1}^{N} h_{s_i}\, h_{e_i}\,e^{j\psi_i }+\Lambda_{2}\,h_{se}  \right |^{2},\label{c7569}    \end{split}\end{align*}
$\Lambda_{1}=\sqrt{\bar{\gamma }_{se}}\left ( \frac{d_{sr}\,  d_{re}}{d_{o}^{2}} \right )^{-\frac{\eta}{2}}
,$ $\Lambda_{2}=\sqrt{\bar{\gamma }_{se}}\left ( \frac{d_{se}}{d_{o}^{2}} \right )^{-\frac{\eta}{2}}
,$ $\bar{\gamma }_{se}=\frac{P_\textup{S}}{\sigma_e^{2}},$ $\Psi_\textup{PE} =\omega_{e}\left | h_{pe}\right |^{2},$ $\omega_{e} =\bar{\gamma }_{e}\,\left ( \frac{d_{sd}}{d_{o}} \right )^{-\eta},$ and $\bar{\gamma }_{e}=\frac{P_\textup{P}}{\sigma_e^{2}}.$ As the phase shifts of the RIS elements are designed based on the legitimate SN link, the resulting phase distributions for each  RIS$\rightarrow $E link $\psi_{i}$ are i.i.d. and uniformly distributed RVs by virtue of \cite{76543}. Thus, $\Psi_\textup{E}$ can be approximated by an exponential RV according to  \cite[Corollary 2]{badiu2019communication} with a parameter $\Lambda_{{E}}=N\, \Lambda_{1}^{2}+\Lambda_{2}^{2}$. Therefore, the  PDF of $\Psi_\textup{E}$ is given by
\begin{equation}
    f_{\Psi_\textup{E}}\left ( \gamma  \right )=\frac{1}{\Lambda_{E}}\,\exp{\left (-\frac{\gamma }{\Lambda_{E} }  \right )}.\label{c76}   
\end{equation}

It is worth mentioning that the SN transmitted power, $P_\textup{S}$, must be under a certain level to limit the interference. Consequently, the interference signal power towards the PR should be constrained as $P_\textup{S}\,\Psi_\textup{P}\leq \mathcal{Q}$, where $\Psi_\textup{P}$ is the summation of the channel power gains from the RIS and S, and $\mathcal{Q}$ is the threshold of interference temperature. Precisely, $\Psi_\textup{P}$ consists of the reflected link $\textup{RIS} \to\textup{PR}$ and the $\textup{S} \to\textup{PR}$ link, while $\mathcal{Q}$ expresses the maximum tolerant interference imposed on the PR. From (\ref{YPP25}), $\Psi_\textup{P}$ can be expressed as
\begin{equation}\begin{split}    \Psi_\textup{P}&=\left |\left ( \frac{d_{sr}  d_{rp}}{d_{o}^{2}} \right )^{-\frac{\eta}{2}}\sum_{i=1}^{N} h_{s_i} h_{p_i}e^{j\theta_i }+  \left ( \frac{d_{sp}}{d_{o}} \right )^{-\frac{\eta}{2}}h_{sp} \right |^{2}.\end{split}    \end{equation}   
Similar to $\Psi_\textup{E}$, $\Psi_\textup{P}$ also can be approximated as an exponential RV with a parameter $\lambda_{\textup{P}}=\left ( \frac{d_{sr}  d_{rp}}{d_{o}^{2}} \right )^{-\eta}N+\left ( \frac{d_{sp}}{d_{o}} \right )^{-\eta}$, where the PDF of $\Psi_\textup{P}$ is given by 
\begin{equation}\begin{split}    f_{\Psi_\textup{P}}\left ( \gamma  \right )=\frac{1}{\lambda_{P}}\,\exp{\left (-\frac{\gamma }{\lambda_{P} }  \right )}. \label{ccp}\end{split}    \end{equation}

\section{Performance Analysis}
\subsection{PN Secrecy Outage Probability}
In this subsection, the  $\textup{SOP}$ of the PN is investigated. The SOP can be expressed as  
\begin{align}   \begin{split}       \textup{SOP}&=\textrm{Pr}\left ( \mathcal{C}_{\textup{S}} < \mathcal{R}_s \right ),\label{c6}    \end{split}\end{align}
where $\mathcal{C}_{\textup{S}}$ is the PN secrecy capacity and $\mathcal{R}_s$ is the PN target secrecy rate. In this regard, $\mathcal{C}_{\textup{S}}$ can be obtained by
\begin{align}   \begin{split}       \mathcal{C}_{\textup{S}}= \big [\mathcal{C}_{\textup{P}}-\mathcal{C}_{\textup{E}} ,0 \big ]^{+},\label{c612}    \end{split}\end{align}
where $\mathcal{C}_{\textup{P}}$ and $\mathcal{C}_{\textup{E}}$ are the PN and the Eav capacities, respectively, and $\left [x,0 \right ]^{+}=\max\left ( x,0 \right )$. Accordingly, $\mathcal{C}_{\textup{P}}$  is given by 
\begin{align}   \begin{split}       \mathcal{C}_{\textup{P}}&=\textrm{log}_2\Big ( 1+\gamma_{\textup{P}}   \Big ),\label{c61}    \end{split}\end{align}
where $\gamma_{\textup{P}} $ is given by

\begin{equation}
    \begin{split}
        \gamma_{\textup{P}} =\frac{P_\textup{P}\left ( \frac{d_{pp}}{d_{o}^{2}} \right )^{-\eta}\left | h_{pp}\right |^{2}}{P_\textup{S}\Psi_\textup{P}+\sigma_p^2}=\frac{P_\textup{P}\left ( \frac{d_{pp}}{d_{o}^{2}} \right )^{-\eta}\left | h_{pp}\right |^{2}}{\mathcal{Q}+\sigma_p^2}=\Phi\left | h_{pp}\right |^{2},
        \label{YPP2}
            \end{split}
\end{equation}
where $\mathcal{Q}=P_\textup{S}\,\Psi_\textup{P},$ $\Phi=\omega_{p}\,\vartheta,$ $\omega_{p}=\frac{P_\textup{P}}{\sigma_p^2}\left ( \frac{d_{pp}}{d_{o}^{2}} \right )^{-\eta},$ and $\vartheta  =\left (\frac{\mathcal{Q}}{\sigma_p^2}+1   \right )^{-1}$. {Antenna
selection approach is employed at the PT to avoid the high hardware complexity while maintaining the diversity and reliability advantages of multiple antennas. More specifically, the importance
of using the antenna selection approach lies in the fact that the power consumption and the
complexity of signal processing overhead are low as compared with other techniques such as
beamforming techniques.} Antenna selection strategy is applied at the PT to maintain multiple antennas' diversity and reliability benefits, while avoiding high hardware complexity. Therefore, the best antenna at PT is selected according to the following criterion
\begin{equation}
    \left | h_{pp} \right |^{2}=\:\underset{n \in \left\{1, ..., N_P \right\} }{\textrm{max}}\left | h_{p_{n}p} \right |^{2}.
\end{equation}
The CDF of ${\gamma_{\textup{P}} }$ is given by 

\begin{equation}
     F_{\gamma_{\textup{P}} }(\gamma)=\sum_{n=0}^{N_P-1}\frac{N_P\binom{N_P-1}{n}}{(-1)^{-n}\left ( n+1 \right )}\left (1-\textup{exp}\left (  \frac{-\gamma\:(n+1)}{\Phi} \right ) \right ).
     \label{s1021}
 \end{equation}
Moreover, $\mathcal{C}_{\textup{E}}$ is given by $\mathcal{C}_{\textup{E}}=\textrm{log}_2\left ( 1+\gamma_{\textup{E}}   \right )$, where $\gamma_{\textup{E}}$ is given in (\ref{YYE}). Now, the SOP can be derived as

\begin{equation}
    \textrm{SOP}_{{\varsigma  }_{}}=\int_{0}^{\infty }F_{{\gamma}_P}(\beta \gamma+\alpha )f_{{\gamma}_{E_{}} }^{\varsigma  }(\gamma )\:d\gamma,
    \label{M018}
\end{equation}
where $\alpha=\beta-1$, $\beta=2^{\mathcal{R}_s}$, and $\varsigma \in\left \{ \textrm{SC},\textrm{MRC} \right \}$. For the SC technique, the PDF of $\gamma_{E_{}}$, $f_{\gamma_{E_{}}}^{\textrm{SC}}(\gamma)$, is given by \cite[eq. (31)]{khoshafa2020improving}
\begin{equation}
\begin{split}
       f_{\gamma_{\textup{E}} }^{\textrm{SC}}(\gamma)=&\sum_{k=0}^{N_{E}-1}\frac{N_{E}(-1)^k\binom{N_{E}-1}{k}}{\omega_{e}\Lambda_{E}}\exp\left (- \frac{\gamma(k+1)}{\omega_{e}}  \right )\\&\times\left (\frac{ 1+\frac{\gamma\left ( k+1 \right )}{\omega_{e}}+\frac{1}{\Lambda_{E}} }{\left (\frac{\gamma\left ( k+1 \right )}{\omega_{e}}+\frac{1}{\Lambda_{E}}  \right )^2}\right ).
        \label{o1042}
        \end{split}
\end{equation}
By plugging (\ref{s1021}) and (\ref{o1042}) into (\ref{M018}), and after simple algebraic manipulations, then with the help of \cite[eq. (3.383.9)]{gradshteyn2014table}, the SOP for SC, $\textrm{SOP}_{\textrm{SC}}$, can be derived as
\begin{equation}
\begin{split}
       \textrm{SOP}_{\textrm{SC}}=&N_P\sum_{n=0}^{N_P-1}\:\frac{(-1)^n\:\binom{N_P-1}{n}}{\left ( n+1 \right )}\left [ 1-\frac{N_{E}}{\omega_{e}\:\Lambda_{E}}\right.\\&\left.\times\sum_{k=0}^{N_{E}-1}\:\frac{(-1)^k\:\binom{N_{E}-1}{k}}{\,\mathcal{H}_{1}\exp\left ( -\left (\mathcal{H}_{3}-\mathcal{H}_{2}  \right ) \right )}\right.\\&\left.\left(\frac{\Lambda_{E}}{\exp\left (- \mathcal{H}_{3} \right )} +\frac{\left ( \mathcal{H}_{4}-\mathcal{H}_{1} \right )}{\mathcal{H}_{1}}\Gamma \left ( 0, \mathcal{H}_{3} \right ) \right ) \right ],
        \label{o10420}
        \end{split}
\end{equation}
where $\mathcal{H}_{1}=\frac{\left ( k+1 \right )}{\omega_{e}}
,$ $\mathcal{H}_{2}=\frac{\alpha \,(n+1)}{\Phi},$ $\mathcal{H}_{3}=\frac{\mathcal{H}_{4}}{\Lambda_{E}\,\mathcal{H}_{1}},$  $\mathcal{H}_{4}=\frac{\beta\left ( n+1 \right )}{\Phi}+\frac{\left ( k+1 \right )}{\omega_{e}}
,$ and $\Gamma (\cdot ,\cdot )$ denotes the upper incomplete gamma function \cite[eq. (8.350.2)]{gradshteyn2014table}. For the MRC technique, the PDF of $\gamma_{E_{}}$, $f_{\gamma_{E_{}}}^{\textrm{MRC}}(\gamma)$, is given by \cite[eq. (37)]{khoshafa2020improving}
\begin{equation}
\begin{split}
       f_{\gamma_{\textup{E}} }^{\textrm{MRC}}(\gamma)=&\frac{\gamma ^{N_E-1}\:\exp\left (\frac{-\gamma }{\omega_{e}}  \right )}{\Gamma(N_E)\:\omega_{e}^{N_E}\Lambda_{E}}\:\sum_{k=0}^{N_E}\frac{\binom{N_E}{k}\Gamma(k+1) }{\left (\frac{\gamma}{\omega_{e}}+\frac{1}{\Lambda_{E}}  \right )^{k+1}}.
       \label{ss0119}
       \end{split}
\end{equation}
By plugging (\ref{s1021}) and (\ref{ss0119}) into (\ref{M018}), and after simple algebraic manipulations, then with the help of \cite[eq. (3.383.4)]{gradshteyn2014table}, the SOP for MRC, $\textrm{SOP}_{\textrm{MRC}}$, can be derived as
\begin{equation}
\begin{split}
       \textrm{SOP}_{\textrm{MRC}}=&\sum_{n=0}^{N_P-1}\frac{N_P\binom{N_P-1}{n}}{(-1)^{-n}\left ( n+1 \right )}\left [ 1-\sum_{k=0}^{N_E}\frac{\binom{N_E}{k}\Gamma(k+1)}{\Lambda_{E}^{{N_E-k}}}\right.\\&\left.\frac{\mathcal{H}_{5}^{-\left ( \frac{N_E-k}{2} \right )}W_{\frac{-N_E-k}{2}, \frac{-N_E+k+1}{2}}\big ( \mathcal{H}_{5} \big )}{\exp\big ( -\left ( 0.5\,\mathcal{H}_{5}-\mathcal{H}_{2} \right ) \big )} \right ],
       \label{ss011901}
       \end{split}
\end{equation}
where $\mathcal{H}_{5}=\left (\frac{\Phi+\beta\,\omega_{e}\,(n+1)}{\Phi\,\Lambda_{E}}  \right )$, and ${W}_{a,b}(\cdot )$ denotes the Whittaker function \cite[eq. (9.220.4)]{gradshteyn2014table}.

\subsection{Asymptotic SOP Analysis}
The PN asymptotic SOP, $\textup{SOP}^{ \infty }$, is studied {when $\omega_{p}\rightarrow \infty$}.  In this scenario, we consider that $ \omega_{p}   >  >    \omega_{e}$. $\textup{SOP}^{ \infty }$ is given by 
\begin{equation} \textup{SOP}^{ \infty }=\left (\mathcal{G}_{a}^{} \overline{ \gamma }_{d} \right )^{-\mathcal{G}_{d}}+\mathcal{O}(\overline{ \gamma }_{d}^{\:-\mathcal{G}_{d}} ),\end{equation}
\noindent
where $\mathcal{G}_d$ is the secrecy diversity order, $\mathcal{G}_a$ is the secrecy array gain, and $\mathcal{O}(.)$ is the higher order terms. In this respect, the $\textup{SOP}^{ \infty }$ can be derived by first obtaining the asymptotic CDF, $F^{ \infty }_{\gamma_{\textup{P}} }(\gamma)$, \cite[eq. (42)]{khoshafa2020physical}. Then, by plugging $F^{ \infty }_{\gamma_{\textup{P}} }(\gamma)$ into (\ref{M018}), and using \cite[eq. (3.382.4)]{gradshteyn2014table} the $\textup{SOP}^{ \infty }$ can be derived. For the SC technique, $\mathcal{G}_{d}^{\textrm{SC}}$ = $N_P$ and $\mathcal{G}_{a
}^{\textrm{SC}}$ is given by
\begin{equation}
    \begin{split}
       \mathcal{G}_{a}^{\textrm{SC}}=&\left [\sum_{k=0}^{N_{E}-1}\sum_{n=0}^{N_{P}}\frac{\binom{N_{E}-1}{k}\binom{N_{P}}{n}\mathcal{Z}_{1}}{(-1)^{-k}\left ( k+1 \right )^{n}}\left (W_{\frac{-n-1}{2}, \frac{-n}{2}}\left ( \frac{1}{\Lambda_{E}} \right ) \right.\right.\\&\left.\left.\times\frac{1}{\sqrt{\Lambda_{E}}}+W_{\frac{-n-2}{2}, \frac{-i+1}{2}}\left (\frac{1}{\Lambda_{E}} \right ) \right )  \right ]^{\frac{-1}{N_P}},
    \end{split}
    \end{equation}
where $\mathcal{Z}_{1}=\frac{N_{E}\,\beta^{n}\,\alpha^{N_P-n}\Gamma \left ( n+1 \right )\,\omega_{e}^{n}   }{\vartheta ^{N_P}\exp\left ( \frac{-1}{2\omega_{e} } \right )}$. For the MRC technique, $\mathcal{G}_{d}^{\textrm{MRC}}$ = $N_P$ and $\mathcal{G}_{a
}^{\textrm{MRC}}$ is given by
\begin{equation}
    \begin{split}
       \mathcal{G}_{a}^{\textrm{MRC}}=&\left [\sum_{k=0}^{N_{E}}\sum_{n=0}^{N_{P}}\:\binom{N_{E}}{k}\binom{N_{P}}{n}\mathcal{Z}_{2}\right.\\&\left.\times W_{\frac{-N_E-k-n}{2}, \frac{-N_E+k-n+1}{2}}\left ( \frac{1}{\omega_{e}} \right )   \right ]^{\frac{-1}{N_P}},
    \end{split}
\end{equation}
where $\mathcal{Z}_{2}=\frac{\beta^{n}\,\alpha^{N_P-n}\Gamma \left ( N_E+n \right )\Gamma \left ( k+1 \right )\,\omega_{e}^{n}   }{\Gamma \left ( N_{E} \right )\vartheta ^{N_P}\exp\left ( \frac{-1}{2\omega_{e} } \right )\,\Lambda_{E}^{\frac{N_E+n-k}{2}} }$.


\subsection{Probability of Non-zero Secrecy Capacity}
In this subsection, the requirement for the presence of non-zero secrecy capacity is investigated. It is worth noting that the non-zero secrecy capacity is achieved when $\gamma_{C}>\gamma_E$. From (\ref{c6}), the PNSC is given by
\begin{equation}
\begin{split}
    \textrm{PNSC}_{{\varsigma  }}=\textrm{Pr}(\mathcal{C}_{\textup{S}}>0)&=\textrm{Pr}\left ( \frac{1+\gamma_{\textup{P}}}{1+\gamma_{\textup{E}}}> 1 \right )
\\&=1-\int_{0}^{\infty }F_{\gamma _{\textup{P}}}(\gamma)f_{\gamma _\textup{E}}^{\varsigma  }(\gamma)\:d\gamma.
    \label{o1}
\end{split}
  \end{equation}
\subsubsection{Eavesdropper's Channel with SC} 
By plugging (\ref{s1021}) and (\ref{o1042}) into (\ref{o1}), and after simple algebraic manipulations, then with the help of \cite[eq. (3.383.9)]{gradshteyn2014table}, the PNSC for SC, $\textrm{PNSC}_{\textrm{SC}}$, can be derived as
\begin{equation}
\begin{split}
       \textrm{PNSC}_{\textrm{SC}}=&1-N_P\sum_{n=0}^{N_P-1}\:\frac{(-1)^n\:\binom{N_P-1}{n}}{\left ( n+1 \right )}\left [ 1-\frac{N_{E}}{\omega_{e}\:\Lambda_{E}}\right.\\&\left.\times\sum_{k=0}^{N_{E}-1}\:\frac{(-1)^k\:\binom{N_{E}-1}{k}}{\,\mathcal{H}_{1}\exp\left ( -\mathcal{H}_{6}  \right )}\right.\\&\left.\times\left(\frac{\Lambda_{E}}{\exp\left (- \mathcal{H}_{6} \right )} +\frac{\left ( \mathcal{H}_{7}-\mathcal{H}_{1} \right )}{\mathcal{H}_{1}}\Gamma \left ( 0, \mathcal{H}_{6} \right ) \right ) \right ],
        \label{o104201}
        \end{split}
\end{equation}
where  $\mathcal{H}_{6}=\frac{\mathcal{H}_{7}}{\Lambda_{E}\,\mathcal{H}_{1}},$  $\mathcal{H}_{7}=\frac{\left ( n+1 \right )}{\Phi}+\frac{\left ( k+1 \right )}{\omega_{e}}
.$

\subsubsection{Eavesdropper's Channel with MRC} 
By plugging (\ref{s1021}) and (\ref{ss0119}) into (\ref{o1}), and after simple algebraic manipulations, then with the help of \cite[eq. (3.383.4)]{gradshteyn2014table}, the PNSC for MRC, $\textrm{PNSC}_{\textrm{MRC}}$, can be derived as
\begin{equation}
\begin{split}
       \textrm{PNSC}_{\textrm{MRC}}=&1-\sum_{n=0}^{N_P-1}\frac{N_P\binom{N_P-1}{n}}{(-1)^{-n}\left ( n+1 \right )}\left [ 1-\sum_{k=0}^{N_E}\frac{\binom{N_E}{k}}{\Lambda_{E}^{{N_E-k}}}\right.\\&\left.\times\frac{\Gamma(k+1)\mathcal{H}_{8}^{-\left ( \frac{N_E-k}{2} \right )}W_{\frac{-N_E-k}{2}, \frac{-N_E+k+1}{2}}\big ( \mathcal{H}_{8} \big )}{\exp\big ( - 0.5\,\mathcal{H}_{8} \big )} \right ],
       \label{ss01190}
       \end{split}
\end{equation}
where $\mathcal{H}_{8}=\left (\frac{\Phi+\omega_{e}\,(n+1)}{\Phi\,\Lambda_{E}}  \right )$, and ${W}_{a,b}(\cdot )$ denotes the Whittaker function \cite[eq. (9.220.4)]{gradshteyn2014table}.

\subsection{SN Outage Probability}
For the SN, the outage probability, $P_{out}$, can be expressed by

\begin{equation}    P_{out}=\textup{Pr}\left ( \gamma_{\textup{D}}\leq 2^{\mathcal{R}_d}-1   \right )=F_{\gamma_{\textup{D}}}(2^{\mathcal{R}_d}-1),    \label{R01}\end{equation}
where $\mathcal{R}_d$ is the SN achievable data rate, and $\gamma_{D}$ is the instantaneous SNR of the CR link. 
Now, by replacing $P_\textup{S}$ with $\frac{\mathcal{Q}}{\Psi_\textup{P}}$ in (\ref{S11}), $\gamma_{\textup{D}} $ can be written as
\begin{equation}    \begin{split}       \gamma_{\textup{D}} =\frac{\mathcal{Q}}{\Psi_\textup{P}\,\sigma_d^2}\left |\left ( \frac{d_{sr}  d_{rd}}{d_{o}^{2}} \right )^{\hspace{-0.2pc}-\frac{\eta}{2}}\hspace{-0.2pc}\sum_{i=1}^{N} h_{s_i} h_{d_i} e^{j\phi_{i}}+  h_{sd}\left ( \frac{d_{sd}}{d_{o}} \right )^{-\frac{\eta}{2}}  \right |^{2},        \label{S1}    \end{split}\end{equation}
which can be rewritten as 
$\gamma_{\textup{D}} =\frac{\Psi_\textup{D}}{\Psi_\textup{P}},
 $ where $\Psi_\textup{D}= \left (\Omega_{1} \sum_{i=1}^{N} \left |h_{s_i}  \right |\, \left |h_{d_i}  \right |+\Omega_{2} \left |h_{sd}  \right | \right )^{2},$ $\Omega_{1} = \frac{{\sqrt{\mathcal{Q}}}}{\sigma_d}\left ( \frac{d_{sr}\,  d_{rd}}{d_{o}^{2}} \right )^{-\frac{\eta}{2}},$ $ \Omega_{2} = \frac{{\sqrt{\mathcal{Q}}}}{\sigma_d}\left ( \frac{d_{sd}}{d_{o}^{2}} \right )^{-\frac{\eta}{2}}
.$ According to the central limit theorem, $\chi _{1}=\sum_{i=1}^{N} \left |h_{s_i}  \right |\, \left |h_{d_i}  \right |$ can be approximated as a Gaussian RV with a mean value $\varepsilon= \frac{N\pi}{4}$ and variance $\sigma^2=N\left ( 1-\frac{\pi}{16} \right )$ \cite{basar2019wireless}. Moreover, $\chi _{2}=\left |h_{sd}  \right |$ is a Rayleigh-distributed RV with a parameter $\delta  $. Thus, the pdfs of $\chi_{1}$ and $\chi_{2}$ are given by
\begin{equation}
    f_{\chi_{1}}(\gamma)=\frac{1}{\sqrt{2\pi\sigma^{2} }}\exp\left ( \frac{-\left ( \gamma -\mu  \right )^{2}}{2\sigma^{2}} \right ),
\end{equation}
and
\begin{equation}
    f_{\chi_{2}}\left (\gamma  \right )=\frac{\gamma }{\delta }\exp\left ( -\frac{\gamma ^{2}}{2\delta } \right ),
\end{equation}
respectively. Thus, $\Psi_\textup{D}$ can be expressed as $\Psi_\textup{D} =\Big (\Omega_{1}  \chi _{1}+\Omega_{2}\,\chi _{2} \Big )^{2},$ leading to the cumulative distribution function (CDF) given by
\cite{yang2020secrecy}
\begin{equation}
    \begin{split}
        &F_{\Psi_\textup{D} }(\gamma )=0.5\left [ \textup{erf}\left ( \frac{(\varphi_1/\varphi_2 )\sqrt{\gamma }-\varepsilon}{\sqrt{2\sigma ^{2}}} \right )+\textup{erf}\left ( \frac{\varepsilon }{\sqrt{2\sigma ^{2}}} \right ) \right ]\\&-\frac{\sqrt{\delta   }}{2\xi_1 }\exp\left ( \frac{-\left (\varphi_1\sqrt{\gamma }-\varphi_2\varepsilon   \right )^{2}}{2\xi _{1}^{2}} \right )\textup{erf}\left ( \frac{\xi _{4}\sqrt{\gamma }-\xi _{5}}{\xi _{1}\xi _{2}} \right )\\&-\frac{\sqrt{\delta   }}{2\xi_1 }\exp\left ( \frac{-\left (\varphi_1\sqrt{\gamma }-\varphi_2\varepsilon   \right )^{2}}{2\xi _{1}^{2}} \right )\textup{erf}\left ( \frac{\xi _{3}\sqrt{\gamma }+\xi _{5}}{\xi _{1}\xi _{2}} \right ),
        \label{Fs}
    \end{split}
\end{equation}
where $\delta$ is a Rayleigh-distributed RV parameter,  $\varepsilon= \frac{N\pi}{4}$, $\sigma^2=N\left ( 1-\frac{\pi}{16} \right )$,   $\varphi_1=\frac{1}{\Omega_{2}}$, $\varphi_2=\frac{\Omega_{1}}{\Omega_{2}}$, $\xi_1=\sqrt{\sigma^{2} \varphi_{2}+\delta    }
$, $\xi_2=\sqrt{2\sigma^{2}\delta  } 
$, $\xi_3=\sigma^{2}\varphi_{1} \varphi_{2} 
$, $\xi_4=\frac{\delta   \varphi_{1}}{\varphi_{2}}$, $\xi_5=\delta\:   \varepsilon $, and $\textup{erf}(\cdot)$ is the error
function \cite[eq. (8.250.1)]{gradshteyn2014table}. $P_{out}$ can be further written mathematically as \cite{papoulis2002probability}
\begin{equation}
    P_{out}=\int_{0}^{\infty }F_{\Psi_\textup{D}}(\gamma\,x)\,f_{\Psi_\textup{P}}(x)\:dx.
    \label{R02}
\end{equation}
However, utilizing (\ref{Fs}) to derive $P_{out}$  is not mathematically tractable. Therefore, the below approximation of the $\textup{erf}(\cdot)$ function is utilized \cite{sadhwani2017tighter}
\begin{equation}
    \textup{erf}\left ( x \right )\approx \left\{\begin{matrix}
1-\sum_{m=1}^{4}\Upsilon_m \exp\left ( -\Theta_m\, x^{2} \right ) & x\geq 0\\ 
 -1+\sum_{m=1}^{4}\Upsilon_m \exp\left ( -\Theta _m\, x^{2} \right )& x< 0,
\end{matrix}\right.
\label{App}
\end{equation}
where $\Theta  = [1, 2, 20/ 3 , 20 /17]$, and $\Upsilon = [1/ 8, 1/4, 1/ 4, 1/ 4]$.

\newpage
\begin{strip}
\begin{equation}
    \begin{split}
      \mathcal{A}_{1}=&\exp\left ( \frac{-b_{2}^{2}}{b_{1}^{2}\lambda_{P} } \right )+\frac{1}{2}\left ( \textup{erf}\left ( \frac{\varepsilon }
{\sqrt{2\sigma^{2} }} \right )-1 \right )+
\sum_{m=1}^{4}\frac{\Upsilon_m\exp\left ( -\frac{a_{2}}{2}  \right )}{4b_{1}^{2}\sqrt{(a_{2}/2)^{3}}\lambda_{P} }\left [  2\sqrt{a_{1}}\left ( \exp\left ( b_{2} \left ( a_{2}-a_{1} b_{2}\right )\right )-2 \right )+\exp\left ( \frac{a_{2}^{2}}{4a_{1}} \right )\right.\\&\left. \sqrt{\pi}\left (  a_{2}-2a_{1} b_{2}\right )\left ( \textup{erfc}\left ( \frac{a_{2}}{2\sqrt{a_{1}}} \right )-\textup{erf}\left ( \frac{a_{2}}{2\sqrt{a_{1}}} \right ) +\textup{erf}\left ( \frac{a_{2}-2a_{1}b_{2}}{2\sqrt{a_{1}}} \right ) \right )\right ],
    \end{split}
\end{equation}
\begin{align}
    \begin{split}
      \mathcal{A}_{2}=&\frac{\exp\left ( -c_{3} \right )\Xi_1}{4}\left [\frac{1}{\sqrt{c_{1}^{3}}}\left ( 2\sqrt{c_{1}}\left ( 2-\exp\left ( \left ( \frac{\xi_{5} }{\xi_{1} \xi_{2} } \right )\left ( c_{2}-c_{1}\left ( \frac{\xi_{5} }{\xi_{1} \xi_{2} } \right ) \right )\right ) \right )-\Xi_2\sqrt{\pi}\exp\left ( \frac{c_{2}^{2}}{4c_{1}} \right ) \left (\textup{erf}\left ( \frac{\Xi_2}{2\sqrt{c_{1}}} \right ) \right.\right.\right.\\&\left.\left.\left.-\textup{erf}\left ( \frac{c_{2}}{2\sqrt{c_{1}}} \right )+\textup{erfc}\left ( \frac{c_{2}}{2\sqrt{c_{1}}} \right )  \right )\right )+
\sum_{m=1}^{4}\frac{\Upsilon_m}{\sqrt{c_{4}^{3}}}\left ( 2\sqrt{c_{4}}\left ( \exp\left ( \left ( \frac{\xi_{5} }{\xi_{1} \xi_{2} } \right )\left ( c_{2}-c_{4}\left ( \frac{\xi_{5} }{\xi_{1} \xi_{2} } \right ) \right )\right )-2\right )\right.\right.\\&\left.\left. +\Xi_3\sqrt{\pi }\exp\left ( \frac{c_{2}^{2}}{4c_{4}} \right )\left ( \textup{erf}\left ( \frac{\Xi_3}{2\sqrt{c_{4}}} \right ) -\textup{erf}\left ( \frac{c_{2}}{2\sqrt{c_{4}}} \right )+\textup{erfc}\left ( \frac{c_{2}}{2\sqrt{c_{4}}} \right ) \right ) \right )  \right ],
\label{Pout}
    \end{split}
\end{align}
\begin{align}
    \begin{split}
      \mathcal{A}_{3}=&\frac{\Xi_{4}}{4}\exp\left ( -\left ( \upsilon _3-\left ( \frac{\xi_{5} }{\xi_{1} \xi_{2} } \right ) \upsilon _2 \right ) \right )\left [\upsilon _{1}^{-\frac{3}{2}} \exp\left ( -\left ( \frac{\xi_{5} }{\xi_{1} \xi_{2} } \right )^2 \upsilon _1 \right )
\left ( 2\sqrt{\upsilon _{1}}-\Xi_{5}\sqrt{\pi}\;\exp\left ( \frac{\Xi_{5}^{2}}{4\upsilon _1} \right )\textup{erf}\left ( \frac{\Xi_{5}}{2\sqrt{\upsilon _1}} \right ) \right )\right.\\&\left.+
\sum_{m=1}^{4}\frac{\Upsilon_m}{\sqrt{\upsilon _{4}^{3}}}\exp\left ( -\left ( \frac{\xi_{5} }{\xi_{1} \xi_{2} } \right )^2 \upsilon _4 \right )\left (-2\sqrt{\upsilon _{4}}+\Xi_{6}\sqrt{\pi}\,\exp\left ( \frac{\Xi_{6}^{2}}{4\upsilon _4} \right )\textup{erf}\left ( \frac{\Xi_{6} }{2\sqrt{\upsilon _4}} \right )   \right )  \right ],
\label{Poutt}
    \end{split}
\end{align}
where
\begin{equation*}
  b_{1}=\frac{ \varphi_
1   \sqrt{\gamma }}{\sqrt{2\sigma ^{2}}},\:\:\:\:  b_{2}=\frac{\varepsilon }{\sqrt{2\sigma ^{2}}},\:\:\:\: a_1=\frac{1}{\lambda _{p}\,b_{1}^{2}}+\Theta _m,\:\:\:\:  a_{2}=\frac{2b_{2}}{\lambda _{p}\,b_{1}^{2}},\:\:\:\:\Xi_{1}=\frac{\xi_{1}\xi_{2}^{2} \sqrt{\delta   } }{\lambda_{P}\xi_{4}^{2}  },\:\:\:\:\Xi_{2}=c_{2}-2c_{1}\left ( \frac{\xi_{5} }{\xi_{1}\xi_{2}} \right ),\:\:\:\:
\Xi_{3}=c_{2}-2c_{4}\left ( \frac{\xi_{5} }{\xi_{1}\xi_{2}} \right ),\end{equation*}
\begin{equation*}
c_{1}=\frac{\varphi_{1}^{2}\xi_{2}^{2}  }{2\xi_{4}^{2}}+\frac{\xi_{1}^{2} \xi_{2}^{2} }{\lambda _{P}\xi_{4}^{2}\gamma },\:\:\:\:
c_{2}=\frac{\xi_{2}}{\xi_{1}\xi_{4}^{2}\gamma }\left ( \varphi_{1}\left ( \varphi _{1}\xi_{5} \gamma -\varepsilon \xi_{4}\gamma \right ) +\frac{2\xi_{1}^{2}\xi_{5}}{\lambda _{P}}\right ),\:\:\:\: c_{3}=\frac{\xi_{5}^{2}}{\lambda _{P}\xi_{4}^{2}}+\frac{1 }{2\xi_{1}^{2}}\left ( \frac{\varphi_{1}\xi_{5}}{\xi_{4}} -\varepsilon\right )^{2},\:\:\:\:c_{4}=c_{1}+\Theta _{m},
\end{equation*}

\begin{equation*}
\upsilon _{1}=\frac{\varphi_{1}^{2}\xi_{2}^{2} \gamma  }{2\xi_{3}^{2}}+\frac{\xi_{1}^{2} \xi_{2}^{2} }{\lambda _{P}\xi_{3}^{2} },\:\:\:\:
\upsilon _{2}=\frac{\xi_{2}}{\xi_{1}\xi_{3}^{2} }\left ( \varphi_{1}\left ( \varphi _{1}\xi_{5} \gamma +\varepsilon \xi_{3}\sqrt{\gamma} \right ) +\frac{2\xi_{1}^{2}\xi_{5}}{\lambda _{P}}\right ),\:\:\:\: \upsilon _{3}=\frac{\xi_{5}^{2}}{\lambda _{P}\xi_{3}^{2}}+\frac{1 }{2\xi_{1}^{2}}\left ( \frac{\varphi_{1}\xi_{5}\sqrt{\gamma }}{\xi_{3}} +\varepsilon\right )^{2},\:\:\:\:\upsilon _{4}=\upsilon _{1}+\Theta _{m},
\end{equation*}
\begin{equation*}
\Xi_{4}=\frac{\xi_{1}\xi_{2}^{2} \sqrt{\delta   } }{\lambda_{P}\xi_{3}^{2}},\:\:\:\:\Xi_{5}=2\upsilon _{1}\left ( \frac{\xi_{5} }{\xi_{1}\xi_{2}} \right )-\upsilon _{2},\:\:\:\:\textup{and}\:\:\Xi_{6}=2\upsilon _{4}\left ( \frac{\xi_{5} }{\xi_{1}\xi_{2}} \right )-\upsilon _{2}.
\end{equation*}
\noindent\rule{\textwidth}{0.4pt}
\end{strip}

\noindent


\begin{figure}[t]
\centering
\includegraphics[width=\linewidth]{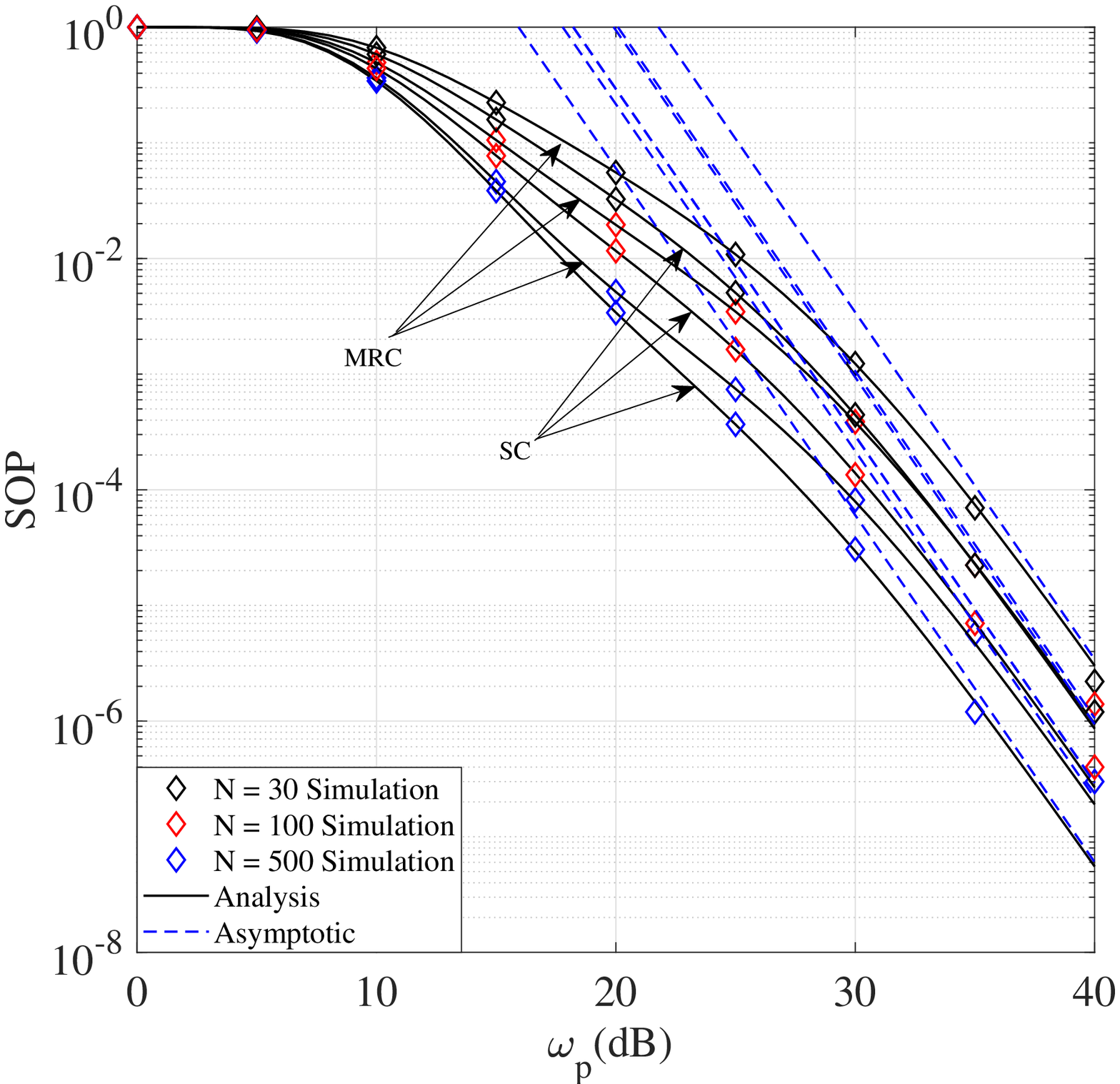}
  \caption{The PN's SOP vs. ${\omega}_{p}$, for different values of the number of reflecting elements, $N$, where $N_P=N_E=3$.}
 \label{fig122}
\end{figure}

\begin{figure}[t]
\centering
\includegraphics[width=\linewidth]{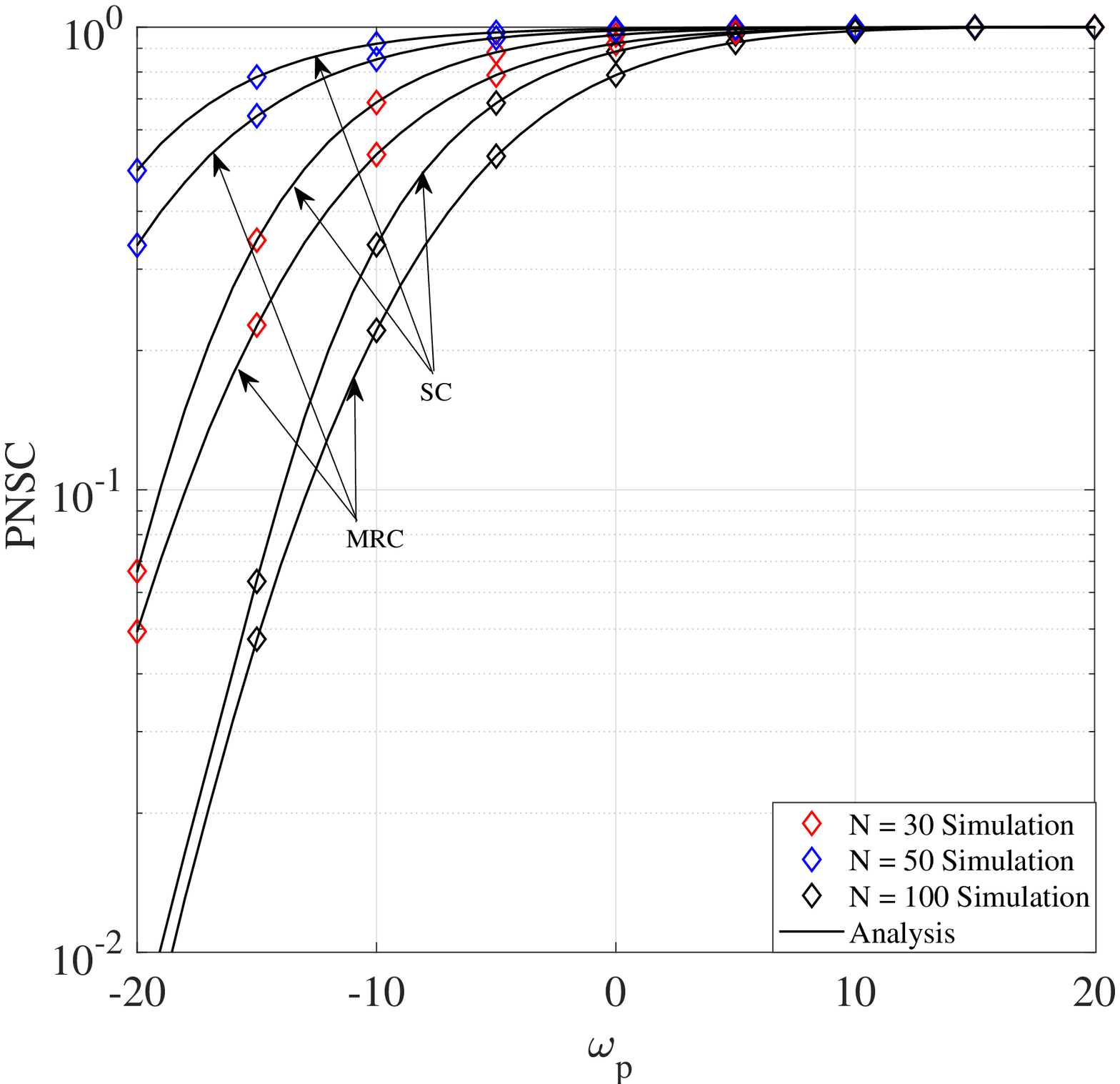}
  \caption{The PN's PNSC vs. ${\omega}_{p}$, for different values of the number of reflecting elements, $N$, where $N_P=N_E=3$.}
 \label{fig13}
\end{figure}

\begin{figure}[t]
\centering
\includegraphics[width=\linewidth]{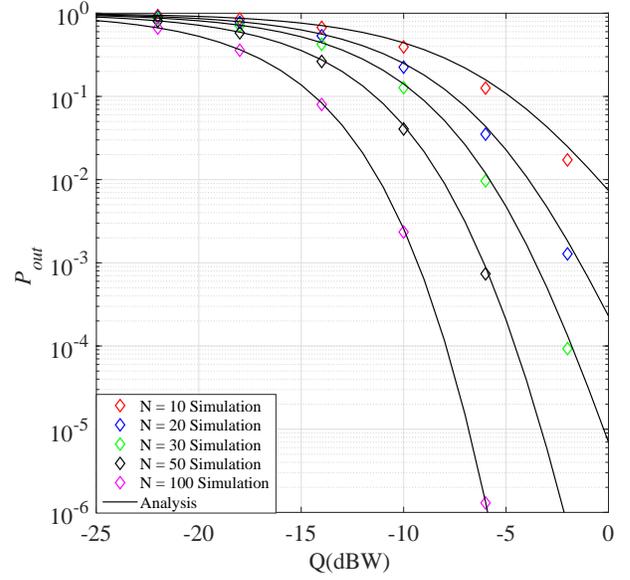}
  \caption{The SN's $P_{out}$ vs. $\mathcal{Q}$, for different values of the number of reflecting elements, $N$, where $\mathcal{R}_d = 1$ b/s/Hz.}
  \label{fig11}
\end{figure}

\begin{figure}[t]
\centering
\includegraphics[width=\linewidth]{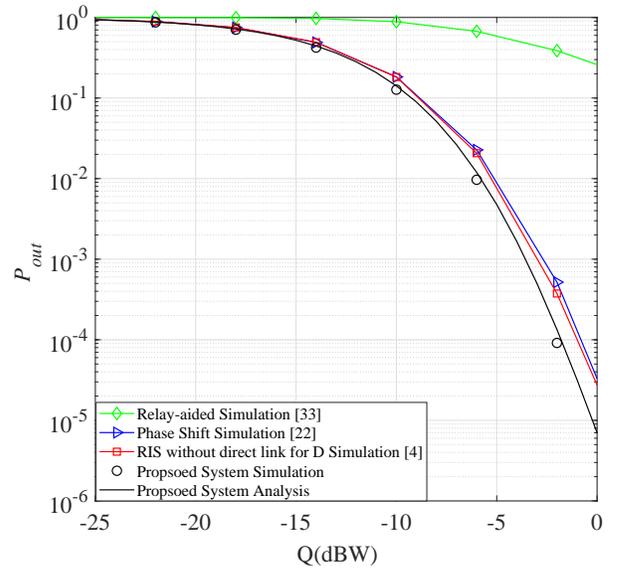}
  \caption{The SN's $P_{out}$ vs. $\mathcal{Q}$, for different
    scenarios, where $N= 30$, $\mathcal{R}_d = 1$ b/s/Hz. }
  \label{fig12}
\end{figure}
\noindent
By substituting (\ref{ccp}) and (\ref{Fs}) to (\ref{R02}) using (\ref{App}), then with the help of \cite[eq. (2.33.1)]{gradshteyn2014table}, $P_{out}$ can be obtained as 
\begin{equation}   P_{out}= \mathcal{A}_{1}-\mathcal{A}_{2}-\mathcal{A}_{3},  \label{I1}\end{equation}
where $\mathcal{A}_{1},$ $\mathcal{A}_{2},$ and $\mathcal{A}_{3}$ are given at the top of this page.
\section{Results and Discussions}
\label{s3}
This section provides numerical and simulation results to confirm
the benefits of applying the RIS technology in the proposed system model. Unless otherwise stated, we set $\mathcal{R}_b$  = 1 b/s/Hz, $\mathcal{Q}$ = 10 dBW,  $\bar{\gamma}_{se}$  = 5 dB,  $\delta$ = 2, and $\mathcal{R}_s$  = 1 b/s/Hz.

In Fig. \ref{fig122}, we investigate the PN secrecy enhancement due to the deployment of the RIS technology. In this respect, the SOP of the PN is evaluated for the SC and MRC techniques versus  $\omega_p$, at Eav, for different values of $N$, where $\omega_e$ = 10 dB. The PN secrecy performance is improved as $N$ increases, showing the effect of the RIS's jamming signals toward Eav. Consequently, the PLS of the PN is increased. Moreover, the $\textrm{SOP}$ is enhanced as $\omega_p$ increases. 
As revealed in our analysis and simulation, improved secrecy performance can be achieved using RIS as a friendly jammer. This is because the wiretapped  signal is degraded at Eav due to the jamming signals generated by the RIS, resulting in a more secure PN transmission. Since the MRC technique produces a higher SNR gain at Eav over the SC technique, the PN secrecy performance is degraded when Eav utilizes the MRC technique, as illustrated in Fig. \ref{fig122}. The asymptotic analyses are included, and perfect agreement with the theoretical results can be seen when $ \omega_p \rightarrow \infty $, confirming the preciseness of the asymptotic expressions. Finally, it is evident that theoretical and simulation results have an excellent match, confirming the exactness of the derived expressions.

Figure \ref{fig13} plots the PNSC versus $\bar{\gamma}_{p}$. It can be noted that the PNSC improves as $\bar{\gamma}_{p}$ increases for a fixed $\bar{\gamma}_e$. Moreover, the PNSC improves with decreasing $\bar{\gamma}_e$. Further, it is also remarkable that the PNSC increases as $N$ increases. Interestingly, secure transmission is guaranteed as $N$ increases. As expected, for the SC technique, the PNSC is lower than that of the MRC technique. Analytical results are also found to match simulation results, validating the accuracy of our analysis.

The SN outage probability, $P_{out},$ is presented in Fig. \ref{fig11}, where the numerical results are provided and compared with the simulated ones. Towards this end, the effect of $N$ on the RIS is evaluated. As shown in this figure, $P_{out}$ of the SN transmission decreases dramatically when $\mathcal{Q}$ increases. With this in mind, the reliability of SN communication increases as $N$ increases. As an illustration, $\mathcal{Q}$ decreases by nearly 4 dB, deploying an RIS technology with $N$ = 20 compared with $N$ = 50 to reach $P_{out}$ = $10^{-2}$.

In Fig. \ref{fig12}, the reliability of the proposed system model is studied and compared with different scenarios. Towards this end, the relay-aided transmission \cite{9060897}, phase shift error, and unavailability of the line of sight between S and D  scenarios are introduced and the results obtained through Monte-Carlo simulations. To evaluate the influence of the discrete phase shifts, simulation results where the phase error is uniformly distributed in $[-\frac{\pi}{4}, \frac{\pi}{4}]$ \cite{9291402}  are provided. Interestingly, the SN's reliability is enhanced by utilizing the RIS in the presence of the S-D link compared to other scenarios. This is due to the fact that the RIS can maximize the received SNR at D and thus improve the channel quality of the SN. It is also noteworthy that simulation and numerical
results match impeccably, verifying the correctness of our analysis. Furthermore, theoretical  results and simulation results agree perfectly, verifying the exactness of our analysis.

\section{Conclusion}
\label{s4}
In this work, the RIS technology is employed to simultaneously assist SN transmission and enhance the PN's secrecy performance in a CR environment. New analytical expressions are provided for the SN's outage probability and the PN's SOP, considering practical combining techniques. The accuracy of the provided expressions is confirmed via extensive Monte-Carlo simulations. Furthermore, the benefits of the proposed system model are verified through numerical and simulation results. 
\balance

\bibliography{mylib}

\begin{thebibliography}{10}
\providecommand{\url}[1]{#1}
\csname url@samestyle\endcsname
\providecommand{\newblock}{\relax}
\providecommand{\bibinfo}[2]{#2}
\providecommand{\BIBentrySTDinterwordspacing}{\spaceskip=0pt\relax}
\providecommand{\BIBentryALTinterwordstretchfactor}{4}
\providecommand{\BIBentryALTinterwordspacing}{\spaceskip=\fontdimen2\font plus
\BIBentryALTinterwordstretchfactor\fontdimen3\font minus
  \fontdimen4\font\relax}
\providecommand{\BIBforeignlanguage}[2]{{%
\expandafter\ifx\csname l@#1\endcsname\relax
\typeout{** WARNING: IEEEtran.bst: No hyphenation pattern has been}%
\typeout{** loaded for the language `#1'. Using the pattern for}%
\typeout{** the default language instead.}%
\else
\language=\csname l@#1\endcsname
\fi
#2}}
\providecommand{\BIBdecl}{\relax}
\BIBdecl

\bibitem{basar2019wireless}
E.~Basar, M.~Di~Renzo, J.~De~Rosny, M.~Debbah, M.-S. Alouini, and R.~Zhang,
  ``Wireless communications through reconfigurable intelligent surfaces,''
  \emph{\textit{IEEE Access}}, vol.~7, pp. 116\,753--116\,773, Jul. 2019.

\bibitem{wu2019towards}
Q.~Wu and R.~Zhang, ``Towards smart and reconfigurable environment: Intelligent
  reflecting surface aided wireless network,'' \emph{\textit{IEEE Commun.
  Mag.}}, vol.~58, no.~1, pp. 106--112, Jan. 2020.

\bibitem{wu2021intelligent}
Q.~Wu, S.~Zhang, B.~Zheng, C.~You, and R.~Zhang, ``Intelligent reflecting
  surface aided wireless communications: A tutorial,'' \emph{\textit{IEEE
  Trans. Commun.}}, vol.~69, no.~5, pp. 3313--3351, May 2021.

\bibitem{RISKhoshafa}
M.~H. Khoshafa, T.~M. Ngatched, and M.~H. Ahmed, ``Reconfigurable intelligent
  surfaces-aided physical layer security enhancement in \uppercase{D2D}
  underlay communications,'' \emph{\textit{IEEE Commun. Lett.}}, vol.~25,
  no.~5, pp. 1443--1447, May 2021.

\bibitem{9530403}
M.~H. Khoshafa, T.~M. Ngatched, M.~H. Ahmed, and A.~R. Ndjiongue, ``Active
  reconfigurable intelligent surfaces-aided wireless communication system,''
  \emph{\textit{IEEE Commun. Lett.}}, vol.~25, no.~11, pp. 3699--3703, Nov.
  2021.

\bibitem{fara2022prototype}
R.~Fara, P.~Ratajczak, D.-T. Phan-Huy, A.~Ourir, M.~Di~Renzo, and J.~De~Rosny,
  ``A prototype of reconfigurable intelligent surface with continuous control
  of the reflection phase,'' \emph{\textit{IEEE Wireless Commun.}}, vol.~29,
  no.~1, pp. 70--77, Feb. 2022.

\bibitem{trichopoulos2022design}
G.~C. Trichopoulos, P.~Theofanopoulos, B.~Kashyap, A.~Shekhawat, A.~Modi,
  T.~Osman, S.~Kumar, A.~Sengar, A.~Chang, and A.~Alkhateeb, ``Design and
  evaluation of reconfigurable intelligent surfaces in real-world
  environment,'' \emph{\textit{IEEE Open J. Commun. Soc.}}, vol.~3, pp.
  462--474, Dec. 2022.

\bibitem{dai2020reconfigurable}
L.~Dai, B.~Wang, M.~Wang, X.~Yang, J.~Tan, S.~Bi, S.~Xu, F.~Yang, Z.~Chen,
  M.~Di~Renzo \emph{et~al.}, ``Reconfigurable intelligent surface-based
  wireless communications: Antenna design, prototyping, and experimental
  results,'' \emph{\textit{IEEE access}}, vol.~8, pp. 45\,913--45\,923, Mar.
  2020.

\bibitem{chorti2022context}
A.~Chorti, A.~N. Barreto, S.~K{\"o}psell, M.~Zoli, M.~Chafii, P.~Sehier,
  G.~Fettweis, and H.~V. Poor, ``Context-aware security for 6\uppercase{G}
  wireless: the role of physical layer security,'' \emph{\textit{IEEE Commun.
  Standards Mag.}}, vol.~6, no.~1, pp. 102--108, Mar. 2022.

\bibitem{basharat2021reconfigurable}
S.~Basharat, S.~A. Hassan, H.~Pervaiz, A.~Mahmood, Z.~Ding, and M.~Gidlund,
  ``Reconfigurable intelligent surfaces: Potentials, applications, and
  challenges for \uppercase{6G} wireless networks,'' \emph{\textit{IEEE
  Wireless Commun.}}, vol.~28, no.~6, pp. 184--191, Dec. 2021.

\bibitem{wyner1975wire}
A.~D. Wyner, ``The wire-tap channel,'' \emph{\textit{Bell sys. tech. j.}},
  vol.~54, no.~8, pp. 1355--1387, Oct. 1975.

\bibitem{9652031}
S.~Fang, G.~Chen, Z.~Abdullah, and Y.~Li, ``Intelligent omni surface-assisted
  secure \uppercase{MIMO} communication networks with artificial noise,''
  \emph{IEEE Commun. Lett.}, 2022.

\bibitem{guan2020intelligent}
X.~Guan, Q.~Wu, and R.~Zhang, ``Intelligent reflecting surface assisted secrecy
  communication: Is artificial noise helpful or not?'' \emph{\textit{IEEE
  Wireless Commun. Lett.}}, vol.~9, no.~6, pp. 778--782, Jun. 2020.

\bibitem{zhang2021improving}
Z.~Zhang, C.~Zhang, C.~Jiang, F.~Jia, J.~Ge, and F.~Gong, ``Improving physical
  layer security for reconfigurable intelligent surface aided \uppercase{NOMA}
  \uppercase{6G} networks,'' \emph{\textit{IEEE Trans. Veh. Technol.}},
  vol.~70, no.~5, pp. 4451--4463, May 2021.

\bibitem{cisco}
\emph{Cisco Annual Internet Report (2018–2023) White Paper}.\hskip 1em plus
  0.5em minus 0.4em\relax {Available [Online]: http://goo.gl/ylTuVx}, Mar.
  2020.

\bibitem{khoshafa2021performance}
M.~H. Khoshafa, J.~M. Moualeu, T.~M. Ngatched, and M.~H. Ahmed, ``On the
  performance of secure underlay cognitive radio networks with energy
  harvesting and dual-antenna selection,'' \emph{\textit{IEEE Commun. Lett.}},
  vol.~25, no.~6, pp. 1815--1819, Jun. 2021.

\bibitem{pan2021reconfigurable}
C.~Pan, H.~Ren, K.~Wang, J.~F. Kolb, M.~Elkashlan, M.~Chen, M.~Di~Renzo,
  Y.~Hao, J.~Wang, A.~L. Swindlehurst \emph{et~al.}, ``Reconfigurable
  intelligent surfaces for \uppercase{6G} systems: Principles, applications,
  and research directions,'' \emph{\textit{IEEE Commun. Mag.}}, vol.~59, no.~6,
  pp. 14--20, Jun. 2021.

\bibitem{zhang2020intelligent}
L.~Zhang, Y.~Wang, W.~Tao, Z.~Jia, T.~Song, and C.~Pan, ``Intelligent
  reflecting surface aided \uppercase{MIMO} cognitive radio systems,''
  \emph{\textit{IEEE Trans. Veh. Technol.}}, vol.~69, no.~10, pp.
  11\,445--11\,457, Oct. 2020.

\bibitem{yuan2020intelligent}
J.~Yuan, Y.-C. Liang, J.~Joung, G.~Feng, and E.~G. Larsson, ``Intelligent
  reflecting surface-assisted cognitive radio system,'' \emph{\textit{IEEE
  Trans. Commun.}}, vol.~69, no.~1, pp. 675--687, Jan. 2021.

\bibitem{shi2022wireless}
E.~Shi, J.~Zhang, S.~Chen, J.~Zheng, Y.~Zhang, D.~W.~K. Ng, and B.~Ai,
  ``Wireless energy transfer in \uppercase{RIS}-aided cell-free massive
  \uppercase{MIMO} systems: Opportunities and challenges,'' \emph{\textit{IEEE
  Commun. Mag.}}, vol.~60, no.~3, pp. 26--32, Mar. 2022.

\bibitem{76543}
H.~Wang, et~al., ``Intelligent reflecting surfaces assisted secure transmission
  without eavesdropper's \uppercase{CSI},'' \emph{\textit{IEEE Signal Process.
  Lett.}}, vol.~27, pp. 1300--1304, Jul. 2020.

\bibitem{9291402}
P.~Xu, et~al., ``Ergodic secrecy rate of \uppercase{RIS}-assisted communication
  systems in the presence of discrete phase shifts and multiple
  eavesdroppers,'' \emph{\textit{IEEE Wireless Commun. Lett.}}, vol.~10, no.~3,
  pp. 629--633, Mar. 2021.

\bibitem{lee2010outage}
J.~Lee, H.~Wang, J.~G. Andrews, and D.~Hong, ``Outage probability of cognitive
  relay networks with interference constraints,'' \emph{\textit{IEEE Trans.
  Wireless Commun.}}, vol.~10, no.~2, pp. 390--395, Feb. 2011.

\bibitem{lei2017secrecy}
H.~Lei, H.~Zhang, I.~S. Ansari, Z.~Ren, G.~Pan, K.~A. Qaraqe, and M.-S.
  Alouini, ``On secrecy outage of relay selection in underlay cognitive radio
  networks over \uppercase{N}akagami-$ m $ fading channels,''
  \emph{\textit{IEEE Trans. Cognitive Commun. Netw.}}, vol.~3, no.~4, pp.
  614--627, Dec. 2017.

\bibitem{bjornson2019intelligent}
E.~Bj{\"o}rnson, {\"O}.~{\"O}zdogan, and E.~G. Larsson, ``Intelligent
  reflecting surface versus decode-and-forward: How large surfaces are needed
  to beat relaying?'' \emph{\textit{IEEE Wireless Commun. Lett.}}, vol.~9,
  no.~2, pp. 244--248, Feb. 2020.

\bibitem{badiu2019communication}
M.-A. Badiu and J.~P. Coon, ``Communication through a large reflecting surface
  with phase errors,'' \emph{\textit{IEEE Wireless Commun. Lett.}}, vol.~9,
  no.~2, pp. 184--188, Feb. 2020.

\bibitem{khoshafa2020improving}
M.~H. Khoshafa, T.~M. Ngatched, M.~H. Ahmed, and A.~Ibrahim, ``Improving
  physical layer security of cellular networks using full-duplex jamming
  relay-aided \uppercase{D2D} communications,'' \emph{\textit{IEEE Access}},
  vol.~8, pp. 53\,575--53\,586, Mar. 2020.

\bibitem{gradshteyn2014table}
I.~S. Gradshteyn and I.~M. Ryzhik, \emph{Table of integrals, series, and
  products}.\hskip 1em plus 0.5em minus 0.4em\relax Academic press, 2014.

\bibitem{khoshafa2020physical}
M.~H. Khoshafa, T.~M. Ngatched, and M.~H. Ahmed, ``On the physical layer
  security of underlay relay-aided device-to-device communications,''
  \emph{\textit{IEEE Trans. Veh. Technol.}}, vol.~69, no.~7, pp. 7609--7621,
  Jul. 2020.

\bibitem{yang2020secrecy}
L.~Yang, J.~Yang, W.~Xie, M.~O. Hasna, T.~Tsiftsis, and M.~Di~Renzo, ``Secrecy
  performance analysis of \uppercase{RIS}-aided wireless communication
  systems,'' \emph{\textit{IEEE Trans. Veh. Technol.}}, Jul. 2020.

\bibitem{papoulis2002probability}
A.~Papoulis and S.~U. Pillai, \emph{\textit{Probability, random variables, and
  stochastic processes}}.\hskip 1em plus 0.5em minus 0.4em\relax Tata
  McGraw-Hill Education, 2002.

\bibitem{sadhwani2017tighter}
D.~Sadhwani, R.~N. Yadav, and S.~Aggarwal, ``Tighter bounds on the gaussian q
  function and its application in \uppercase{N}akagami-m fading channel,''
  \emph{\textit{IEEE Wireless Commun. Lett.}}, vol.~6, no.~5, pp. 574--577,
  Oct. 2017.

\bibitem{9060897}
M.~H. Khoshafa, T.~M.~N. Ngatched, M.~H. Ahmed, and A.~Ibrahim, ``Secure
  transmission in wiretap channels using full-duplex relay-aided
  \uppercase{D2D} communications with outdated \uppercase{CSI},''
  \emph{\textit{IEEE Wireless Commun. Lett.}}, vol.~9, no.~8, pp. 1216--1220,
  Aug. 2020.

\end{thebibliography}
\bibliographystyle{IEEEtran}

\end{document}